\documentclass[aps,prb,preprint,groupedaddress,showpacs,tightenlines]{revtex4}

\usepackage{graphicx}

\begin{document}

\title{Sound modes broadening in quasicrystals.}

\author{M. de Boissieu}
\affiliation 
{Lab. de Thermodynamique et Physico - Chimie Metallurgiques,
UMR CNRS 5614, INPG, UJF, BP75, 38402,  St. Martin d'H\'{e}res Cedex, France}

\author{R. Currat} 
\affiliation {Laue-Langevin Institute, F-38042, Grenoble, France }

\author{S. Francoual} 
\affiliation 
{Lab. de Thermodynamique et Physico - Chimie Metallurgiques,
UMR CNRS 5614, INPG, UJF, BP75, 38402,  St. Martin d'H\'{e}res Cedex, France}
\affiliation {Laue-Langevin Institute, F-38042, Grenoble, France }

\author{E. Kats} \affiliation {Laue-Langevin Institute, F-38042, Grenoble, France } \affiliation {L. D.
Landau Institute for Theoretical Physics, RAS, 117940 GSP-1, Moscow, Russia}

\date{\today}

\begin{abstract}
We propose a simple phenomenological model to
analyze 
vibrational characteristics of quasicrystals (QCs). The interpretation of the
obtained recently data
is based on the existence of almost dispersionless 
optical modes most probably related to the 
specific clusters which constitute the characteristic building blocks of any
QC structure. 
We generalize to QCs the well - known  
Akhiezer
mechanism  
(responsible for the absorption of sound
even in an ideal crystal), which in our case is related to
a ''long wave'' disturbance of the quasicrystalline optical modes
by the propagating sound modes.
At higher wave vectors strong hybridization of acoustic and optical modes
takes place, and it leads to a more steep broadening dependence
on wave vectors, and besides the excitation can no longer be
described as a
single acoustic mode with a well defined wave vector.
We show that the observed sound - mode attenuation
behavior can be consistently
described by these scenarios without invoking 
additional mechanisms.

\end{abstract}

 
\maketitle

\section{Introduction}
\label{int}

In most many body systems there is a fairly sharp distinction between 
elementary excitations (quasi - particles) and the collective modes
involving a coherent mixture of quasi - particles. In a harmonic dielectric
crystal these elementary excitations are phonons, having the linear dispersion
law. At finite temperatures $T$ the great majority of phonons
have energies at most of the order of $T$ due to the nature of the Bose - Einstein
distribution, and the thermodynamic properties of crystals are determined
almost entirely by these excitations. The wave vector associated with a 
phonon in a regular lattice is
only defined modulo a reciprocal lattice vector, a feature which plays a crucial role
in the dynamical properties of crystals. In this respect phonons in crystals are quite
different from sound waves in liquids which carry true momentum.
Each atom in the crystalline lattice executes only a finite amplitude motion (the
oscillation about the lattice site). Evidently the mean momentum of such a motion is
identically zero, and therefore the phonon flux associated with
the energy flux in a crystal is not accompanied by a transfer of mass.

QC systems are quite different from conventional solid crystals.
QCs are built roughly speaking according to the following principles:
being homogeneous
like crystalline alloys QCs nevertheless posses specific structural elements,
or clusters (see \cite{rev1}, \cite{mon1}, \cite{mon2}).
The resulting construction has no translational symmetry but
has long-range order.
In spite of the fact that QC systems do not show lattice periodicity, the 
above statement on zero mean momentum of phonons should be true for QCs also. Thus one must
admit
that phonons in QC systems carry not momentum but only quasi - momentum,
which is defined, unlike in conventional crystals, modulo a dense set of reciprocal 
vectors \cite{KC96}.
Both systems (crystals and QCs) in this respect are different from
liquids, where
the phonon momentum is the actual momentum and the phonon
flux does involve a transfer of mass, since in a sufficiently
long time any atom or molecule can reach any point in the volume.

Thus one could say that i-QCs are quasicrystalline on the near atomic scale, and isotropic
solids at macroscopic scales. Somewhere there ought to be a crossover
from the local arrangement to macroscopic isotropy. If this structural crossover
is sufficiently abrupt, it might also mark in reciprocal space the upper
frequency end of the isotropic acoustic branches.
Besides, because QC systems do not show lattice periodicity,
the concept of Brillouin zone and its use for characterizing elementary excitations
like phonons, strictly speaking is not adequate \cite{JA00} (see also \cite{KL03}, \cite{SR03}).
On the other hand, at least at first glimpse, it can be argued that because
in the long - wavelength limit 
the phonons can be described as sound waves propagating through an elastic continuum,
there should be little difference between QCs and crystalline materials in this
regime where the phonons are insensitive to microscopic structure.
Nevertheless, experimental results are fairly different in the two types of systems.
The following correlation is characteristic of classical mono-crystals:
the higher the quality of the diffraction pattern of a mono-crystal,
the lower the line-widths of the phonons. The correlation
reflects the wave nature of phonons: the better the conditions
for the propagation, the smaller the scattering of the Bloch-type
waves. Since QCs do not posses lattice periodicity  vibrational excitations
QCs are
intrinsically not Bloch-type
eigen-modes and therefore this correlation does not hold.

The same kind of problem has been discussed in the literature
on electronic excitations in QCs.
Some unusual electronic properties (e.g. a very low value of the
electrical conductivity, a negative temperature coefficient of the resistivity,
and so on) as it was shown recently \cite{RM97} are the consequences
of the spectral properties of quasiperiodic Hamiltonians. It is common
knowledge now that Anderson localization \cite{AN84} is a second - order
phase transition between eigenstates that are spatially localized
and those that are delocalized, and there are well defined conditions
at which the localization - delocalization transition occurs.
Many numerical investigations of quasiperiodic Hamiltonians reveal
that their eigenstates are critical, i.e. they are characterized
by a power-law decay of the amplitudes (see \cite{JA96}, \cite{FM96}).
The existence of the critical electronic states
can be understood as related to the self-similarity of the QC
structures: a given atomic cluster is repeated within a distance 
of the order of its size, and therefore an eigenstate from any
cluster can easily tunnel to the next one which has an identical
form. Thus electronic properties of QCs are critical, at least
from the point of view of the second order Anderson localization - delocalization
transition.

However 
this analogy can not 
be bluntly 
applied to the study of vibrational modes
in QCs. There are two principal differences between the Anderson electron
and the vibrational problems, which are our main concern here.
Firstly, when the system is mechanically stable, there are no
negative eigenvalues (unlike in the electron case). Secondly in the 3d
vibrational case, there are always at least 3 zero - frequency (in the long
wavelength limit) Goldstone modes that can not be localized. Therefore,
since the modes at the lower energy bound of the spectrum must 
have extended character, one can
expect something similar to critical states (and to localization -
delocalization transitions) only near the relatively high - frequency
band edge.

The issue (elementary vibrational excitations in QCs)
was intensively studied recently, 
and
inelastic neutron
scattering data (see e.g. \cite{DB99} - \cite{SK02}, 
\cite{rev1}, \cite{HK93} 
and the monographs \cite{mon1},
\cite{mon2}) reveal
in the long wavelength limit well defined acoustic modes.
The experimental quantity investigated in these experiments, the so - called
scattering function $S({\bf Q}, \omega)$, is measured close to
a Bragg reflection located at a certain reciprocal vector ${\bf G}$ in momentum space,
and it probes phonons with frequency $\omega $ and 
wave vector ${\bf q} = {\bf G} - {\bf Q}$.
To measure the largest signal one should measure 
$S({\bf Q}, \omega )$ 
at
high momentum transfer and close to a strong Bragg peak.
Neutron scattering investigations were performed for different icosahedral 
\cite{DB99} - \cite{CO98},
and decagonal \cite{SK02} 
structures 
and it turned out that in spite of the fact that, structurally,
decagonal and icosahedral QCs are rather different,
and that their physical properties (e.g. elastic moduli) might be 
also very different,
they exhibit many common vibrational features.
The purpose of our paper is to propose a simple phenomenological model to
try to obtain as much theoretical
information concerning vibrations in QCs as is possible on a phenomenological
level without specific microscopic treatment.
The rest of the paper is organized as follows:
in Section \ref{exp} we give a short outline of neutron data
related to vibrational mode studies in i-QCs $Al Pd Mn$ and $Zn Mg Y$.
Section \ref{mod} is devoted to the description of our phenomenological model.
Finally, Section \ref{con} deals with miscellaneous subjects related
to the vibrational dynamics in QCs.
 
\section{Experimental overview}
\label{exp}

The excitation spectrum of different QCs has been studied by inelastic neutron scattering on
single grain samples.
In spite of the often contradictory results of experimental
investigations (see e.g. discussions in \cite{SF02}, \cite{SL02}),
a few conclusions listed below seem inescapable. 
As explained above, it is only in the acoustic regime 
that a one to one
correspondence can be made between the observed signal 
and a phonon, i.e. that the phonon 
wave vector can be defined. 
Generally, results are presented in an 'extended' zone scheme, with a strong
Bragg peak acting as the zone centre (for crystallographic details see 
e.g. \cite{CS86}). As proposed by
Niizeki,
\cite{NI89}, \cite{NA90} important pseudo Brillouin zone boundaries
(PZB) can be defined. 
They correspond to strong Fourier components of the atomic
density from which acoustic modes are Bragg
reflected. The positions of these pseudo zone 
boundaries obviously depend on the details of the
atomic structure. 
Moreover, because of the quasi periodicity, several successive PZBs are stacked around 
each
strong zone centre. 

In the following we will focus on two icosahedral phases for which detailed studies
have been carried out: $i - Al Pd Mn$ and $i - Zn Mg Y$ . These two icosahedral phases have a very
different atomic structure. 
Nevertheless their excitation spectra share a lot of common features. 
Their dynamical
response can be separated into two well defined regimes: 
the acoustic regime for wave vectors smaller than
$ 0.6\, \AA ^{-1}$, and, for larger wave vectors, a regime in which the dynamical response is
characterised by a broad
band of dispersionless optic-like modes. 
In the acoustic regime, excitations are generally resolution
limited for $q$ smaller than $0.3\, \AA ^{-1}$, and then broaden. 

The way in which the broadening
develops strongly
depends on the atomic structure of the icosahedral phase: 
for instance it is much more rapid in the
$i - Al Pd Mn$ phase than in the $i - Zn Mg Y$ phase. 
The optic like spectrum generally consists of 3 or 4
broad 'bands' (a few $meV$ wide) around 7, 12, 16 and 24 $meV$. 
Although no gap opening could be observed at
the PZBs,
the energy of the lower optical 
modes roughly corresponds to the intersection of the acoustic branch
with the PZBs. 

Figure 1 summarises results obtained for transverse modes propagating along a 2-fold axis
and polarised in a 2-fold direction. 
The upper part is for the $i-AlPdMn$ phase and the lower one for the
$i-ZnMgY$ one. The transverse acoustic phonon dispersion relation is given by the full circles. The dashed
line is the corresponding sound velocity as extracted from ultrasonic measurements.  The open
circles
correspond to the width of the $TA$ modes. The other open symbols correspond to the different broad optic
'bands', whose width is of the order $4\, meV$. 
For both phases the crossover between the acoustic and, say, hybridized 
regime is fairly abrupt: for the $i-AlPdMn$ phase it shows up as a rapid broadening of the $TA$ 
acoustic
mode,
whereas for the $i-ZnMgY$ there is a coupling with an 
optic like excitation localised at $10\, meV$. In both
cases the $TA$ limit is given by $q = 0.6\, \AA ^{-1}$. 
Above this value the $TA$ acoustic mode mixes up with
optical excitations and the observed signal 
can no longer be described as a single excitation. 

The broadening of
$TA$ excitations also depends on the structure: 
it is about two times larger in the $i-AlPdMn$ phase than in
the $i-ZnMgY$ one. From the local slope of the $TA$ dispersion 
relation and width of the excitations it is
also interesting to compute the mean free path of $TA$ phonons at the upper limit of the acoustic
regime. We
find a mean free path of 
$12\, \AA $ and $24\, \AA $ for the $i-AlPdMn$ and $i-ZnMgY$ phases, respectively. 
Structural X-ray studies \cite{SF98} 
have shown one special geometric aspect
of all QC structures, namely, atomic clusters, with a characteristic diameter $D_{cl} = 10\,
\AA $, which
are part of
the building
blocks of QCs and their
close approximants. 
Summarising the experimental dynamics study, we can thus say that the
crossover between the acoustic and 
mixed regime occurs for a phonon wavelength equal to $D_{cl}$. The mean free
path of the $TA$ phonon is equal to $D_{cl}$ and $2D_{cl}$ for 
the 
$i-AlPdMn$ 
and $i-ZnMgY$ phases, respectively.

Line broadening is a nuisance in some circumstances, while in our case of QCs
it provides valued physical information. We show in Fig. 2
dependences of line width 
for longitudinal (the upper panel) modes  
propagating along a 5-fold axis 
and for the transverse modes (the lower panel)
propagating along a 2-fold axis 
in the $i-ZnMgY$ phase, and the width for the transverse mode
propagating along a 2-fold axis for 
$i-AlPdMn$ is shown in Fig. 3.
The instrumental width 
for longitudinal modes (about 2 mev)
is larger 
than for transverse (about 1 mev). To gain further
insight into the nature of broadening we show in Figs. 2 and 3
theoretical fitting by $q^2$ dependence for the longitudinal waves and by $q^4$
for the transverse modes. We will see in the next section that these
dependences correspond to broadenings dominated respectively by Akhiezer  
and by resonance hybridization sound absorption.

It is worth noting that very similar results are obtained for
physically and structurally very different systems like i-QC $Cd Yb$ or
decagonal - QC
$Al Ni Co$ (see \cite{DB99} - \cite{SK02}).
Even numerical values for the acoustic line-width cross-over are
not very different, however, to be specific 
we restrict ourselves
in this paper to the i-QCs $Al Pd Mn$ and $Zn Mg Y$ only for which
we have more detailed and reliable data due to the high quality
of the samples.

\section{Phenomenological Akhiezer model}
\label{mod}

Whereas the knowledge on the sound propagation
in ordered crystals or disordered glass-like structures is now
rapidly improving (see e.g., \cite{CF03}, \cite{RF03}, and references therein), 
little is known about the behavior in systems with non-periodic long-range order,
such as QCs.
The general objective of this study is to determine the structural
and physical mechanisms associated with phonon line broadening in order
to identify those features of QCs which differentiate them from conventional
crystalline
systems.
Since both features (characteristic clusters and almost dispersionless
optical modes) are interrelated and practically ubiquitous for 
all QCs 
it is natural
to look for a model based
on these widely recognized specific
features.
The primary aim of this section is to present such a physical model
that reproduces available experimental data on vibrational eigenmodes
in QC. In particular we are going to consider general
phenomenological processes leading
to sound absorption and dispersion, as the detailed microscopic
mechanisms governing
these processes are not fully understood yet and have been heavily
debated in the literature \cite{SF02}, \cite{SL02}. 

The occurrence of many broad optical, almost dispersionless modes has been, at least qualitatively,
established and their origin can be easily understood. Indeed, as is well known \cite{AM76},
the number of optical modes in a crystal is related to the number of
atoms in the unit cell. Since in QC systems the unit cell size is strictly speaking
infinite, 
it is not surprising to have in a certain energy window a dense
set of optical modes.
Moreover due to natural energetic restrictions the energy window where all the
optical modes are confined is a relatively narrow one,
which unavoidably leads to small mode
dispersion.
We shall thus include from the very beginning these modes into the 
model. On the other hand the dispersionless character of the optic modes
means that the corresponding excitations see an almost uniform medium,
the strain of which is slowly modulated by the sound wave.
Due to the anharmonicity of the system the strain of the sound
field causes a change of the optic mode frequencies.
Despite the apparent analytic intractability of the problem, the various
approaches suggest that all the complications can be subsumed by a simple
phenomenological form.

In order for the discussion to proceed smoothly, we start first with a simple
argument. We generalize to QCs the well - known  
Akhiezer mechanism 
of sound absorption (see \cite{AK39} and also \cite{BD60}). 
In the case (we will be interested in below) when the frequency
of the sound waves is much smaller than all relevant inverse relaxation times,
the sound vibrations can be considered as a certain external
field which modulates the QC structure, and therefore the optical mode
frequencies
\begin{eqnarray}
\label{a1}
\omega _0^{(\alpha , a)} = \omega _0^{(\alpha , a)} (1 + \gamma _{i j}^{(\alpha , a)} u_{i j})
\, ,
\end{eqnarray}
where $\alpha $ denotes the polarization
of the optic mode $a$, $\gamma _{i j}^{(\alpha , a)}$ is the second rank
tensor which characterizes the QC and depends on the direction of the
optic mode $(\alpha \, , a)$ propagation. It can be viewed 
as the generalized Gruneisen constant tensor (see \cite{AK39} and also
\cite{BD60}).

To simplify the matter (though it is not a crucial point) we assume that
$\gamma $ does not depend on the frequency
$\omega $. It can be shown using standard thermodynamic
relations \cite{LL80} that in this case the optical mode
frequency modulation does not disturb the thermodynamic equilibrium
within this branch, thus it is still proper to attribute a temperature $T$
to it
\begin{eqnarray}
\label{a2}
\left (\frac{\Delta T}{T}\right )^{(\alpha , a)} \propto
\left (\frac{\Delta \omega _0}{\omega _0}\right )^{(\alpha , a)}
\, .
\end{eqnarray}
This relation (\ref{a2})
is valid if during the strain deformation collisions with other branches
can be neglected.
However since for different modes the values of $\gamma $ can be different
(even could be negative or positive), after a strain deformation each branch
has in general a different temperature (some branches might be cooled
down, while others are heated up).

It might be useful first to separate all optical modes into two groups 
with different average temperatures but comparable specific heats.
The average relative temperature
difference between the groups, and according to (\ref{a1}),
(\ref{a2}), the average generalized Gruneisen coefficient 
$\hat {\gamma } _{\alpha \, \rm {av}}$ 
can be then easily calculated. Indeed one should bear in mind that although the
modes contribute additively to the free energy, the Gruneisen coefficients
are determined by a ratio of derivatives, i.e. a ratio of sums
$$
\hat {\gamma } _{\alpha \, \rm {av}} 
= \frac{\sum _{a} \hat {\gamma } ^{(\alpha \, a)} C_a}{\sum _{a} C_a}
\, ,
$$
where $C_a$ is the specific heat associated to the mode $a$.
 
If a sound wave propagates through a system, a periodic temperature
difference will be set up between the two groups. Therefore
in a certain characteristic relaxation time $\tau $, heat exchange takes
place between them, leading to entropy production,
and absorption of the sound wave. The absorption coefficient
per unit length $\alpha _\alpha $ (in $cm^{-1}$) is determined
by the entropy production rate and reads
\cite{LL86}
\begin{eqnarray}
\label{a3}
\alpha _\alpha  = \frac{C_\alpha  T \gamma _{\alpha , \rm {av}}^2}{\rho c _\alpha ^3} \frac{\omega ^2
\tau }{1 +
\omega ^2 \tau ^2}
\, ,
\end{eqnarray}
where $C_\alpha $ is the heat capacity at constant strain corresponding to each sound mode
polarization (constant volume for longitudinal waves and constant shear for transversal
ones),
$\rho $ is the mass
density, and $c_\alpha \equiv (c_l \, , c_t) $ is the longitudinal or transverse 
sound mode velocity.

It seems reasonable to assume that the energy exchange 
between the two groups takes place by means of
phonon - phonon scattering processes. The relevant point is that the same
kind of processes determine the heat conductivity \cite{LL81}
$\kappa $
\begin{eqnarray}
\label{a4}
\kappa  = \frac{1}{3} C_v c_l^2  \tau
\, ,
\end{eqnarray}
where $C_v$ is the constant-volume specific heat, and $c_l$ is the longitudinal
mode velocity.

The expressions can be brought into more compact forms
by taking into account 
that at high enough temperatures the relaxation time $\tau $
is much shorter than the sound period.
Comparing (\ref{a3}) and (\ref{a4})
we end up with the following coefficients for the longitudinal
($\alpha _l$) and transverse ($\alpha _t$) sound absorption
\begin{eqnarray}
\label{a6}
\alpha _l\simeq  \frac{\kappa \gamma ^2_{l, av}}{\rho c_l^5} T \omega ^2 
\, ,
\end{eqnarray}
and
\begin{eqnarray}
\label{a12}
\alpha _t \simeq  \frac{\kappa \gamma ^2_{t, av}}{\rho c_t^3 c_l^2}\frac{C_t}{C_v}
T \omega ^2 
\, ,
\end{eqnarray}
where $C_t$ is the heat capacity at constant shear strain.

For the benefit of the skeptical reader, some comments about
the applicability  of the mechanism described in this
section to transverse modes, seem in order.
Indeed conventional wisdom claims that the Akhiezer mechanism
cannot be applied directly to
transverse sound modes, since only dilatation or compression deformations may provide non -
uniform temperature variations and thermo-diffusive dissipation.
To resolve the contradiction we should recall that the building
blocks of all QC structures are atomic clusters.
These clusters are not mere geometrical constructions
but real physical entities responsible for specific
features in the QC vibrational spectrum
(e.g., responsible for localized modes).
Hence, one expects acoustic vibrations to induce stress inhomogeneities 
around each cluster, with a resulting strain field that will
consist of a superposition of a dilation and of a shear.

Therefore in this case the internal thermoelastic
dissipation occurs even for the pure transverse vibrations.
It is reminiscent of a phenomenon known in classical elasticity theory \cite{LO44} 
whereby a spherical inclusion placed in a vibrating medium containing only pure 
shears, introduces local strains
which have dilation $\Delta (r, \theta , \phi )$ ($r$, $\theta $, and $\phi $
are spherical coordinates) proportional
to the shear deformation $s$
at large distances from the inclusion
\begin{eqnarray}
\label{i1}
\Delta = - s \left (\frac{5\mu }{9 \lambda + 14 \mu }\right ) \left (\frac{D_{cl}}{r}\right )^2
\sin ^2\theta \cos \phi \sin \phi
\, ,
\end{eqnarray}
where, for an isotropic material, 
$\mu $ and $\lambda $ are the only two non-zero 
elastic constants.
To find Akhiezer line broadening following from the dilation
(\ref{i1}) one should calculate and average the corresponding
Gruneisen coefficient. For one spherical inclusion in a bulk
isotropic material the total effect will be zero due to angular
integration. But of course this picture of one isolated cluster in a vibrating medium
is not adequate for QC clusters. A complete account of
the strains induced by the sound modes in real QC structures
requires considerable computational work, that is beyond the scope
of our purely phenomenological model. 
Thus we use in (\ref{a12}) phenomenologically the average transverse 
Gruneisen coefficient $\gamma _{t , av}$.
However from simple arguments presented above concerning the 
expression (\ref{i1}) we can come to the conclusion that due to  
the cluster-like nature of all QC structures, the 
Akhiezer broadening is expected
to be less efficient for the transverse phonons than for the longitudinal
ones.
Therefore at least for the former case we have to look for another
mechanisms contributing to transverse mode broadening in QCs.

QC clusters play also a crucial role for sound mode  
hybridization in QCs. 
Indeed, if we consider the vibrations of an isolated cluster of size $D_{cl}$, 
then the
high-frequency modes with $\lambda \ll D_{cl}$ 
will not be affected by the change in the boundary
conditions at
scale $D_{cl}$, involved in partial disconnecting the cluster from the rest of the system. 
On the other hand, the
low-frequency modes for which $\lambda \gg D_{cl}$ 
will disappear from the spectrum. 
Very qualitatively one
can then
think of such a material as a dense packing of connected clusters.
Phonon scattering from fluctuations in the local velocity of sound $c^2(r)$, depends on the
variance parameter 
$$ 
\nu (\lambda ) \equiv \frac{\langle c^4(r)\rangle _\lambda }{\langle c^2(r)\rangle ^2_\lambda } - 1
\simeq \nu (a) \left (\frac{a}{\lambda }\right )^3 \, ,
$$ 
where $\lambda = 2\pi /q$ is the phonon wavelength, and $a$ is the
scale at which elementary fluctuations occur (i.e. for QCs $a \simeq D_{cl}$).
Let us have $\nu (a) \gg 1$.
Since $c^2(r)$ is always positive, the only way this can occur is by having a small
number of "hard" regions for which $c^2$ is larger than the average, which dominate 
the variance, and a
much larger number of "soft" regions. If the "hard" scatterers (of size $D_{cl}$) are isolated, the only
propagating modes are those of the "soft" background medium. The scatterers act as rigid inclusions in the
medium. Roughly, the "hard" scatterers simply increase the effective elastic constant by a factor
proportional to the small volume factor they occupy.
The above 
argument 
neglects the local modes centred around the "hard" scatterers. For a single scatterer
they would be localized and have a high frequency. When they are dilute and uncorrelated, 
so that they couple
weakly, they can only form an relatively high-frequency optical band. On the other hand, when they are
correlated and coupled sufficiently strongly, they can dominate the elastic properties and the
low-frequency velocity and damping of sound.
As a rule of thumb we can estimate the characteristic frequency for
mode hybridization as $c_\alpha /D_{cl}$. 
In this region phonon wavefunctions are repetitively localized
on the clusters and this recurrent localization comes from the
fact that, unlike for periodic crystals, it is not possible
for the wavelength to be commensurate simultaneously
with all inter-cluster distances in a quasiperiodic structure.
Evidently this cluster vibration mode is not an exact eigenstate.
It interacts with other vibration modes of similar energy and
as well with acoustic phonons. This interaction affects
both the cluster vibration mode and the phonons, which are
for the same reason also not exact eigenstates.
In other words due to 
the above 
described 
mode - hybridization the excitations will be broadened and shifted.

Physically these cluster vibrations are caused by a local deviation of
the force constant matrix and atomic mass from the average values.
The hybridization phonon broadening $\Gamma $ for a phonon with wave vector
$\bf k$, polarization $j$, and frequency $\omega _0$ depends on a ''concentration''
(i.e. density of states $g(\bf {k})$) of these cluster vibrations and is proportional
to the phonon - cluster mode scattering matrix $\hat t$
 \begin{eqnarray}
\label{i11}
\Gamma (\omega _0(\bf {k}, j)) = g(\bf {k}) \frac{1}{2} \frac{\Im \langle \bf {k}, j |\hat t| \bf {k} , j
\rangle }{\omega _0 (\bf {k} , j)}
\, .
\end{eqnarray}
To obtain total hybridization broadening this expression (\ref{i11})
has to be averaged and integrated over all cluster vibrations with
frequencies approximately equal to $\omega _0$. If the structure
of the mode does not strongly depend on its frequency, and their
distribution is smooth on a scale of their width, $\Gamma $ will
have $k$-dependence coming mainly from the density of states factor
in (\ref{i11}). Using the arguments borrowed from
the standard glass  theory \cite{PH80} we can conclude that $g \propto k^4$.

\section{Discussion and Conclusion}
\label{con}

Let us summarize now the experimental results and the theoretical
interpretation presented in this paper.
The physical essence of the model is as follows.
If the wavelength $\lambda $ of the sound is larger than the
characteristic cluster size $D_{cl}$ we can assume
that each cluster is subject to a uniformly distributed
stress. However due to not spherical cluster shape and boundary
conditions at their surfaces, the resulting deformation
is not uniform. It varies considerably over the dimension
of a cluster (not over sound wavelength as in a uniform bulk
body). Hence sound broadening due to thermal conduction
(Akhiezer mechanism) is essential for the case.
To be specific this scenario can be rationalized as 
sound modes (mainly longitudinal but as we have seen
also transverse) 
somehow perturb the manifold of optic
modes
or QC cluster 
vibrations  
(the characteristic features of QC materials taken into account by the
model). These perturbations in terms of
Gruneisen parameters are reduced to temperature variations, and the 
latter ones lead to thermo-diffusional
relaxational sound absorption. 
It might at first sight seem that the model contains no QC-specific features: but
in fact all QC properties are hidden
in the material parameter values entering the formula. 

The discussed above (in the previous section \ref{mod}) second broadening mechanism
- resonance hybridization of sound and cluster vibration modes -
can compete with the Akhiezer absorption even for the longitudinal
modes in the relatively high $q$-region. All the more it is true
for the transverse waves where the Akhiezer broadening is expected
to be less efficient. Fortunately they do have different $q$-dependencies
that allow us to disentangle them analysing experimental data.
To span a wide range of possibilities let us discuss first
the Akhiezer mechanism.
To compare the expressions with neutron data it is more
convenient to use:

(i) the absorption coefficient $\Gamma [s^{-1}] = \alpha [cm^{-1}] c_\alpha $;
 
(ii) the wave vector $q$ instead of $\omega $. 

Thus we come from (\ref{a6}) to
\begin{eqnarray}
\label{d1}
\Gamma = \frac{\kappa \gamma _{av}^2 T}{\rho
c_\alpha ^3} q^2
\, . 
\end{eqnarray}

The proposed model can be checked against experimental 
data along three lines: the wave vector dependence
of the acoustic mode
broadening which should be proportional to $q^2$, 
its temperature variation and its order of magnitude.
 
The $q^2$ increase of the acoustic phonon width was generally observed for longitudinal acoustic
excitations.
For $TA$ modes the width increase seems to be more abrupt, 
(for a wave vector larger than 
$0.3\, \AA ^{-1}$) 
and goes rather as a $q^4$ law as shown in Figs. 2 and 3.
Presented by the solid line in Fig. 1 
a $q^2$ fit to the observed 
widths, illustrates this statement, and besides, can be used
to extract some quantitative information on the magnitudes
of $q^2$ and $q^4$ parameters.
As expected, the coefficient of 
the fit is twice as large in the $i-AlPdMn$ phase as in the
$i-ZnMgY$ one (9.8 and 4.9 $meV/ \AA ^{-2}$). In the $i-ZnMgY$, 
we also found that the $LA$ mode broadens twice as fast as the $TA$ modes.
More accurate inspection of the data presented in Fig. 1 clearly shows
that for a region close to the mixed regime (i.e. to $q \simeq 0.6 \, \AA $) 
the broadening is more
rapid than a
$q^2$ law. It can be attributed to Rayleigh scattering
arising from local fluctuation in sound velocity due to either
mass density or elastic constant fluctuations (in own turn related
to the cluster structure of all QCs) or to acoustic - optic mode
hybridization. In mathematical form both phenomena
lead to $q^4$ broadening \cite{PH80}, though of course the Rayleigh
scattering seems generally too weak in homogeneous materials
like QCs.

The next test concerns temperature dependences. 
Since 
at high enough temperatures $\kappa \propto 1/T$, in 
the present approximation 
the effect of the temperature dependent factors cancels, and therefore
the absorption due to Akhiezer mechanism can only weakly depend on 
temperature. 
It is worth noting also that the very possibility to
characterize the system by only one Gruneisen function is
based on the assumption that 
the dependence of the energy
levels 
on the  
volume (or more generally
on the deformations $u_{ik}$) 
is expressible in terms of a single characteristic
energy. This assumption is correct for Einstein or Debye models,
partly also for conventional (with relatively small elementary cell) 
crystalline solids but not
evident
at all for QC systems where a number of different contributions to the free
energy can control the thermodynamic properties (see discussion in \cite{SF02}, \cite{SL02}).

Thus in this approximation (namely, in the
high temperature regime where $T > \Theta_D$, $\Theta _D$ being the Debye temperature, 
400 - 500 K, within
the
classical Debye approach valid in this region, $\kappa \propto 1/T$ (all other parameters determining
$\Gamma $ do not depend noticeably on $T$) and
$\Gamma $ is temperature
independent in agreement with experimental data. 
Of course it is only an approximation and we are aware that
QCs are not classical Debye insulators, and besides we are not in the regime $T \gg \Theta _D $,
but it identifies correctly the characteristic scales in the problem.
A detailed temperature dependence of the lattice
dynamics has only been carried out in the $i-AlPdMn$ phase. At $T = 1050 K$ the slope of the acoustic
transverse acoustic mode 
displays only 
a $10\% $ decrease as compared to room temperature and
the broadening of the modes did not show any significant variation.
This is thus in agreement with the model since the Debye temperature of the $i-AlPdMn$ phase is about $
500\, K $(see e.g. \cite{SF02}).

Quantitative comparison of our model predictions  with experimental data is more difficult since
there are only scarce data available for the model input parameters and their
temperature dependences (there are no systematic measurements of all needed material parameters,
and experimental data are still not very accurate). 
Thus from here on in this 
section we shall not attempt to maintain numerical accuracy, but only indicate
the
form of the answer.                                                    
Our model can be used to estimate the $q^2$ coefficient in the phonon
line broadening.
Namely, from 
\cite{SF02}, \cite{SL02}, \cite{GS00}:
$$ 
\kappa \simeq 1\frac{W}{m K} = 10^5\, \frac{erg}{s cm K}  
\, ,
$$ 
$$ \rho = 5\, g/cm^3 \, ;
\quad \gamma _{av} \simeq 1 
\, ,
$$ 
and, besides, $T \simeq 300\, K $. 
Using also the literature (\cite{BB93}, \cite{BB95})
data for i-QC materials  
($Zn Mg Y$
$c_l = 4.8\cdot 10^5\, cm/s $, $c_t = 3.1\cdot 10^5\, cm/s$), and neutron measurements
(\cite{BB93}, \cite{BB95} and our data shown in Figs. 1 - 3),
one can find that at $ q \simeq 0.5 \,\AA ^{-1}$, 
for the longitudinal phonons, $\Gamma _l \simeq 1.5\, meV $, and for
the transverse modes, depending on propagation directions $\Gamma _t \simeq  1\, meV$ ;
for $ Al Pd Mn$
($c_l = 6.3\cdot 10^5\, cm/s $, $c_t = 3.5\cdot 10^5\, cm/s$), at the same 
$ q \simeq 0.5 \, \AA ^{-1}$, $\Gamma _t \simeq 2\, meV$ (no 
data for longitudinal mode
line-widths).
Putting all values together we find 
for the $q^2$ coefficient about $10\, meV/\AA ^2$ in reasonable agreement with the
experimental value quoted above.
 
One more interesting line of thought is to 
apply our results to estimate the anisotropic Gruneisen parameters,
introduced in (\ref{a1}). 
The observed difference in
width increase rate between transverse and 
longitudinal excitations means that a different Gruneisen
parameter
should be used in each case. 
Comparing experimental data for longitudinal
and transverse modes in the same material (for the moment such data are
available only for $i- Zn Mg Y$) we can conclude that the volume Gruneisen
parameter entering the longitudinal phonon broadening (\ref{a6}) should be
four 
times larger than the shear anisotropic Gruneisen coefficient which determines 
the transverse phonon
line broadening (\ref{a12}).

These Gruneisen parameters can be seen as a phenomenological way of taking into account the
interaction between $LA$ or $TA$ modes and optical branches.
Various phonon interaction processes result
when anharmonic terms of the third (and higher) order in the displacements are
taken
into account. The first anharmonic term corresponds to the decay of one phonon into two
or to the coalescence of two colliding phonons into one. For conventional
crystals, usually the main contribution comes from processes within the same optical
branch. For QCs, the dispersionless character of  
the optical branches could lead to a situation
where the conditions for phonon decay or recombination are satisfied
above a finite threshold wave vector (corresponding
to the crossing with the lowest frequency optical branch), and 
not for a discrete set of wave vectors but for a whole spherical shell in $q$ space.

The coefficient at $q^4$ for the resonance hybridization broadening,
is a model dependent quantity. However, it can be always presented in the
following form \cite{PH80}
\begin{eqnarray}
\label{xnew}
\alpha _{\alpha , {\rm res}} = \Lambda _\alpha \frac{\omega }{c_\alpha }\left (\frac{\omega 
}{\omega _{cl}}\right )^3
\, ,
\end{eqnarray}
where $\Lambda _\alpha $ is a model and mode polarization dependent
coefficient, and $\omega _{cl}$ is a characteristic frequency for
cluster vibrations.
To illustrate this issue we show in Figs. 2 and 3 $q$-dependences of line width 
for longitudinal and transversal modes
in the $i-ZnMgY$ and 
$i-AlPdMn$ QCs.
These results are clearly indicating that for the longitudinal modes
the broadening are governed mainly by the Akhiezer mechanism, whereas
for the transversal waves $q^4$ fitting leads to a reasonable agreement
with the data. Moreover the coefficient at $q^4$ found from such a fitting
is conformed with (\ref{xnew}) describing 
resonance hybridization sound absorption.

Of course the imperfect knowledge of the parameters in a large
temperature interval makes our predictions only qualitatively or semi-quantitatively
correct, and at this stage, a number of open questions must be stressed. For example no
forbidden gaps have been observed
experimentally. There is also some  
inconsistency between thermodynamic and inelastic neutron scattering
data \cite{SF02},
\cite{SL02}. A possible origin of this inconsistency may be related to
contributions, say to the specific heat, from the cluster vibrations.
It is worth noting, however, that all scattering experiments measure a dynamic structure factor
$S(q, \omega )$, which is the Fourier transform of the displacement correlation
function, and therefore does not carry direct information about
propagation or non-propagation of modes.
But to put these speculations on a firm footing,
further
experimental and theoretical efforts are required.
Nevertheless for a model with such a small amount of physical input,
our results show quite good agreement with experiments
(the order of magnitudes together with its evolution with $T$ and
$q$). 
In addition we can give some qualitative predictions concerning Gruneisen parameters
and anharmonicity in different QC materials.
It is worth noting in this respect one very recent
theoretical prediction \cite{ZG02} that Gruneisen parameters should diverge
close to a quantum critical point. In a certain sense QC phonon and electron states
are critical over the whole region of the QC state stability.
However, it should be noted that dynamics of QC
is still a developing field and much of the excitement arises from the
possibility of discovering novel physics beyond say the
classical paradigms discussed here.

As this paper was being written for publication we became aware of parallel
efforts \cite{CF03}, \cite{RF03} to investigate the crossover from
propagating to strongly attenuated acoustic modes in densified silica glasses.
Although physically the two systems (QCs and glasses) are very different,
acoustic mode broadening reported in these papers 
and its interpretation
(Akhiezer attenuation due to optical mode perturbations and acoustic -
optic modes scattering and hybridization) are quite similar to our results.
Of course the nature of these optical modes is different in each case:
in glasses it is related to disorder (in silica glasses it corresponds
to librations of $Si O_4$ tetrahedra), in QCs the optic modes are related
to cluster vibrations or, in other words, to 
the translational or orientational frustrations intrinsic to all QCs.

\acknowledgements  

One of us (E.K.) acknowledges support from 
INTAS (under No. 01-0105) Grants.
We thank T. Janssen for thought-provoking discussions.


\newpage

\centerline{Figure caption}

Fig. 1

Dispersion relation measured around the strong 2-fold Bragg
reflection with $N/M$ indices equal to 52/84 and propagating along a 2-fold axis in the $i-AlPdMn$ 
(the upper figure)
and $i-ZnMgY$ phase (the lower figure). Results are presented in an extended zone scheme, the pseudo
Brillouin zone
boundaries are shown as vertical dashed lines. The transverse acoustic mode positions are shown as full
circles and their width as open one. The dashed line is the linear dispersion relation as deduced from
ultrasonic measurements and the solid line is a $q^2$ fit to the width increase. The other open symbols
correspond to dispersionless optic like excitations, whose width is of the order $4 meV$. 

Fig. 2

Line width 
for longitudinal (the upper figure) modes  
propagating along a 5-fold axis (experimental data and $q^2$-fitting)
and for the transversal modes (the lower figure)
propagating along a 2-fold axis (experimental data and $q^4$-fitting)
in the
$i-ZnMgY$ 
phase.

Fig. 3

Line width 
for the transversal mode 
propagating along a 2-fold axis (experimental data and $q^4$-fitting)
in the
$i-Al Pd Mn$ phase.


\begin{references} 
\bibitem{rev1} M.Quilichini, T.Janssen, Rev. Mod. Phys., {\bf 69},
277 (1997).
\bibitem{mon1} R.Bellissent, M. de Boissieu, G.Coddens in Physical properties
of quasicrystals, Z.M. Stadnik, ed., Springer (1999).
\bibitem{mon2} C.Janot, Quasicrystals: a Primer, Oxford, Oxford Science
(1992).
\bibitem{KC96} P.A.Kalugin, M.A.Chernikov, A.Bianchi, H.R.Ott,
Phys. Rev. B., {\bf 53}, 14145 (1996).
\bibitem{JA00} T.Janssen, Ferroelectrics, {\bf 236}, 157 (2000).
\bibitem{KL03} M.Kleman, Eur. Phys. J., B, {\bf 31}, 315 (2003).
\bibitem{SR03} V.V.Savkin, A.N.Rubtsov, T.Janssen, {\bf 31}, 525 (2003).
\bibitem{RM97} S.Roche, D.Mayou, Phys. Rev. Lett., {\bf 79}, 2518
(1997).
\bibitem{AN84} P.W.Anderson, Basic Notions of Condensed Matter Physics,
Adison and Wesley, Frontier in Physics (1984).
\bibitem{JA96} C.Janot, Phys. Rev. B, {\bf 53}, 181 (1996).
\bibitem{FM96} T.Fujiwara, T.Mitsui, S.Yamamoto, Phys. Rev. B, {\bf 53}, 2910 (1986).
\bibitem{DB99}
F.Dugain, M.de Boissieu, K.Shibata, R.Currat, T.J.Sato, A.R.Kortan, J.B.Suck,
K.Hradil, F.Frey, A.P.Tsai, Eur. Phys. J. B, {\bf 7}, 513 (1999).
\bibitem{BB93} M. de Boissieu, M.Boudard, R.Bellissent, M.Quilichini,
B.Hennion, R.Currat, A.I.Goldman, C.Janot, J.Phys.: Condens. Matter, {\bf 5},
4945 (1993).
\bibitem{BB95} M.Boudard, M.de Boissieu, S.Kycia, A.I.Goldman, B.Hennion,
R.Bellissent, M.Quilichini, R.Currat, C.Janot, J.Phys.: Condens. Matter,
{\bf 7}, 7299 (1995). 
\bibitem{CO98} M.A.Chernikov, H.R.Ott, A.Bianchi, A.Migliori, T.W.Darling,
Phys. Rev. Lett., {\bf 80}, 321 (1998). 
\bibitem{SK02} K.Shibata, R.Currat, M. de Boissieu, T.J.Sato,
H.Takakura, A.P.Tsai, J.Phys.: Condens. Matter, {\bf 14}, 1847 (2002).
\bibitem{HK93} J.Hafner, M.Krajci, J.Phys.: Condens. Matter,
{\bf 5}, 2489 (1993).
\bibitem{SF02}
C.A.Swenson, I.R.Fisher, N.E.Anderson, Jr., P.C.Canfield, A.Migliori, Phys. Rev. B,
{\bf 65}, 184206 (2002).
\bibitem{SL02}
C.A.Swenson, T.A.Lograsso, A.R.Ross, N.E.Anderson, Jr., Phys. Rev. B,
{\bf 66}, 184206 (2002).
\bibitem{CS86} J.W.Cahn, D.Shechtman, D.Gratias, J.Mat. Res., {\bf 1}, 13 (1986).
\bibitem{NI89} K.Niizeki, J.Phys. A: Math. Gen., {\bf 22}, 4295 (1989).
\bibitem{NA90} K.Niizeki, T.Akamatsu, J.Phys.: Cond. Matter., {\bf 2}, 2759 (1990).
\bibitem{SF98} W.Steurer, F.Frey, Phase Transitions, {\bf 67}, 319 (1998). 
\bibitem{CF03} E.Courtens, M.Foret, B.Hehlen, B.Ruffl\'e, R.Vacher, J.Phys.:
Condens. Mattter, {\bf 15}, 1281 (2003).
\bibitem{RF03} B.Ruffl\'e, M.Foret, E.Courtens, R.Vacher, G.Monaco,
Phys. Rev. Lett., {\bf 90}, 095502 (2003).
\bibitem{AM76} N.W. Ashcroft, N.D.Mermin, Solid State Physics,
Holt, Rinehart and Winston, New York (1976).  
\bibitem{AK39} A.Akhiezer, J.Phys. (USSR), {\bf 1}, 277 (1939).
\bibitem{BD60} H.E.B\"ommel, K.Dransfeld, Phys. Rev., {\bf 117}, 1245 (1960).
\bibitem{LL80} L.D.Landau, E.M.Lifshits, Statistical Physics, Part 1,
Pergamon Press, New York (1980). 
\bibitem{LL86} L.D.Landau, E.M.Lifshits, Theory of elasticity, Pergamon
Press, New York (1986).                                                                                         
\bibitem{LL81} L.D.Landau, E.M.Lifshits, Physical Kinetics (Course of
Theoretical Physics, volume 10), Pergamon Press, New York (1981).
\bibitem{LO44} A.E.H.Love, A treatise on the mathematical theory of elasticity,
Chapter XI, Dover, Oxford (1944).
\bibitem{PH80} W.Phillips (ed.), Amorphous Solids - Low Temperature Properties,
Springer, New York (1980).
\bibitem{GS00} K.Gianno, A.V. Sologubenko, M.A.Chernikov, H.R.Ott, I.R.Fisher, P.C.Canfield
Phys. Rev. B, {\bf 62}, 292 (2000).
\bibitem{ZG02} L.Zhu, M.Garst, A.Rosch, Q.Si, Phys. Rev. Lett., {\bf 91},
066404 (2003).


\end{references}
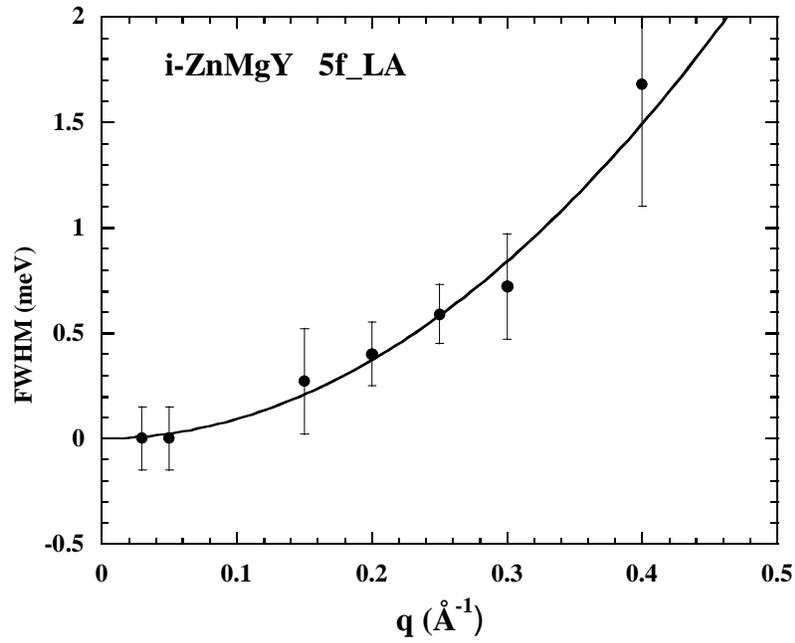
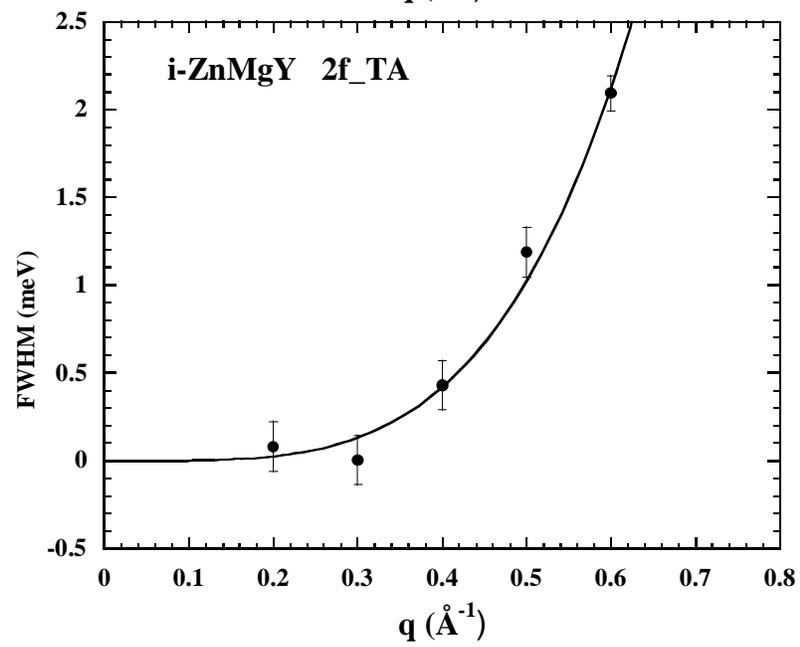
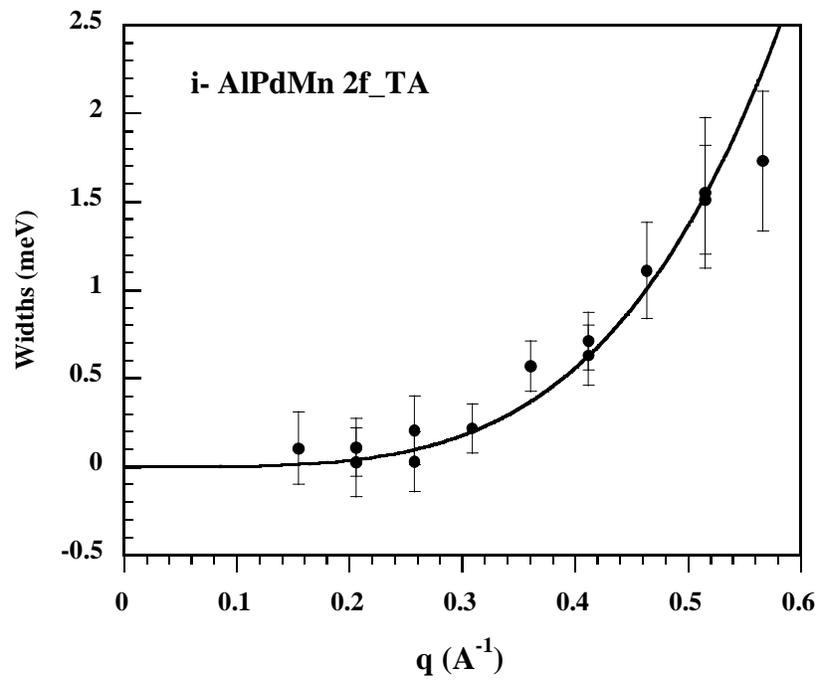
\end{document}